\documentclass[12pt]{iopart}
\usepackage{epsfig,graphicx}

\newcommand{\diag}{{\rm diag\,}}

\newcommand{\trg}{{\rm trg\,}}
\newcommand{\detg}{{\rm detg\,}}

\begin{document}

\title[Norm--dependent Random Matrix Ensembles in
         External Field and Supersymmetry] {Norm--dependent Random
         Matrix Ensembles in External Field and Supersymmetry}

\author{Thomas Guhr}
\address{Matematisk Fysik, LTH, Lunds Universitet,
                 Box 118, 22100 Lund, Sweden}

\begin{abstract}
The class of norm--dependent Random Matrix Ensembles is studied in the
presence of an external field. The probability density in those
ensembles depends on the trace of the squared random matrices, but is
otherwise arbitrary.  An exact mapping to superspace is performed. A
transformation formula is derived which gives the probability density
in superspace as a single integral over the probability density in
ordinary space. This is done for orthogonal, unitary and symplectic
symmetry. In the case of unitary symmetry, some explicit results for
the correlation functions are derived. 
\end{abstract}

\pacs{05.45.Mt, 05.30.-d, 02.30.Px}



\section{Introduction}
\label{sec1}

Supersymmetry is a prominent and widely used tool in studying
disordered systems and systems that can be modeled by random matrices,
see Refs.~\cite{EFE83,EFE97,VWZ,GMGW,Haake,TGenc}. The method was
developed for Gaussian probability densities, a review and a
discussion of the mathematical justification was recently given in
Ref.~\cite{ZIR04}. This restriction to Gaussian probability densities
is no shortcoming if one is exclusively interested in calculating
correlations on the local scale of the mean level spacing. This is due
to local universality~\cite{BZ93a,BZ93b}. Probability densities which
do not introduce scales competing with the mean level spacing yield
correlations which are one the local scale identical to the ones
resulting from Gaussian probability densities, see a review in
Ref.~\cite{GMGW}.  When studying matrix models in high--energy physics
one is not interested in the local scale. Another universality in the
leading asymptotics of the matrix dimension was found~\cite{AJM90} for
the correlation functions on large scales.

Nevertheless, restriction to Gaussian probability densities does not
always suffice. First, the one--point functions obviously depend on
the specific form of the probability densities, because they are not
measured on local scales. Such level densities are important, for
example for applications in high energy physics~\cite{BIPZ78}, but
also in finance~\cite{LCBP99}. Second, to distinguish certain
directions in matrix space, one adds an external field to the random
matrix, and one often averages over the matrices representing the
external field.  The local correlations now change and depend
sensitively on the root--mean square matrix element of the external
field divided by the local mean level spacing. Examples are the
crossover transitions, see the review in Refs.~\cite{GMGW,Haake}. In
such a situation, scales competing with the local mean level spacing
might occur which can lead to a deviation from universal features,
such that the crossover transitions would differ for different
probability densities. Third, non--Gaussian probability densities and
their non--universal features on special scales have always been of
interest in conceptual studies and for considerations in general
statistical mechanics, we mention the bound--trace and the
fixed--trace ensembles~\cite{Mehta} and the recently introduced
ensembles deriving from a non--extensive entropy
principle~\cite{TVT04,BBP04}.

In the present contribution, we show that the supersymmetry method can
be extended to random matrix models with non--Gaussian probability
densities. In the context of universality, asymptotic considerations
for infinite level number have already been combined with
supersymmetric techniques for non--Gaussian probability densities in
Ref.~\cite{HaWe}.  Here, however, we aim at an exact discussion. We
focus on the large class of norm--dependent random matrix ensembles
which depend through an arbitrary functional form on the trace of the
squared random matrices. Recently, a general construction of these
ensembles was given in Ref.~\cite{MK05}. For the reasons just
mentioned, we include an external field. We have two goals. First, we
want to deliver the conceptually important insight that supersymmetry
is by no means restricted to Gaussian probability densities. Second,
we want to provide a series of explicit and practically relevant
formulae for the correlation functions. Here, the application of
supersymmetry yields particularly handy results in the presence of an
external field.

The question whether or not norm--dependent ensembles can be
formulated exactly in a supersymmetric framework was also discussed by
F.~Kalisch. Although his approach was quite different from the one to
be presented here, it would have been likely to produce equivalent
results for the case without external field. Unfortunately, F.~Kalisch
left academia and his findings are unpublished.

The paper is organized as follows.  We formulate the problem in
Section~\ref{sec2}, thereby also introducing our notations and
conventions. The supersymmetric representation of the norm--dependent
ensembles is constructed in Section~\ref{sec3}.  In
Section~\ref{sec4}, we discuss a series of examples. Explicit results
for the correlation functions are given in Section~\ref{sec5}. We
summarize and conclude in Section~\ref{sec6}.

\section{Formulation of the Problem}
\label{sec2}

In Section~\ref{sec21}, we set up the generating function in the
presence of an external field. As we need to refer to the Gaussian
case, we briefly sketch it in Section~\ref{sec22}. We discuss
norm--dependent ensembles and pose the problem in Section~\ref{sec23}.

\subsection{Generating Function in the Presence of an External Field}
\label{sec21}

The three symmetry classes of $N\times N$ random matrices $H$ are
labeled by the Dyson index $\beta$. In the orthogonal class, $H$ is
real symmetric ($\beta=1$) and in the unitary class, $H$ is Hermitean
($\beta=2$). In the symplectic class, $H$ is self--dual ($\beta=4$)
and the entries of $H$ are $2\times 2$ quaternions.  The eigenvalues
of $H$ are doubly degenerate in the symplectic class. We notice that
such a matrix has
\begin{equation}
\mu = N +\frac{\beta}{2}N(N-1)
\label{deg}
\end{equation}
independent matrix elements. The quantity $\mu$ is often referred to
as the number of degrees of freedom.  A normalized probability density
function $P^{(\beta)}(H)$ defines --- together with the symmetry class
--- the random matrix ensemble.  We add a fixed external field
represented by a matrix $H_0$ which, without loss of generality, can
be assumed to be diagonal. In the symplectic class, it has dimension
$2N\times 2N$. Thus, we are interested in a system described by
$H_0+\alpha H$ where $\alpha$ measures the relative strength.  The $k$
level correlation function $R_k^{(\beta)}(x_1,\ldots,x_k,\alpha,H_0)$
is the probability density to find $k$ eigenvalues of $H_0+\alpha H$
at positions $x_1,\ldots,x_k$. The correlation functions
$\widehat{R}_k^{(\beta)}(x_1,\ldots,x_k,\alpha,H_0)$ are technically
easier to handle. They include real and imaginary parts of the
propagator, while the $R_k^{(\beta)}(x_1,\ldots,x_k,\alpha,H_0)$ are
only the correlations of the imaginary parts. The latter can easily be
constructed from the former. We use the conventions of
Refs.~\cite{TG,Guh96,GUH4}.  For arbitrary $P^{(\beta)}(H)$, the
correlation function can be written as the derivative
\begin{equation}
\widehat{R}_k^{(\beta)}(x_1,\ldots,x_k,\alpha,H_0) = 
                                      \frac{1}{(2\pi)^k} \frac{\partial^k}{\prod_{p=1}^k\partial J_p}
                                      Z_k^{(\beta)}(x+J) \Bigg|_{J_p=0}
\label{corr}
\end{equation}
of a generating function
\begin{equation}
Z_k^{(\beta)}(x+J) = \int d[H] \, P^{(\beta)}(H) \prod_{p=1}^k \left( 
\frac{\det(H_0+\alpha H - x_p - J_p)}{\det(H_0+\alpha H - x_p + J_p)}
\right)^\gamma 
\label{gen}
\end{equation}
with respect to source variables $J_p , \ p=1,\ldots,k$.  Here, we
define $\gamma=1$ if $\beta=1,2$ and $\gamma=2$ if $\beta=4$, moreover
we introduce the diagonal matrices $x=\diag(x_1,x_1,\ldots,x_k,x_k)$
and $x=\diag(-J_1,+J_1,\ldots,-J_k,+J_k)$.  The volume element $d[H]$
is simply the product of the differentials of all independent matrix
elements.  For complex variables, we use the differentials of real and
imaginary part.

\subsection{Gaussian Random Matrix Ensembles}
\label{sec22}

In the Gaussian case, the normalized probability density function with
variance $2v^2/\beta$ reads
\begin{equation}
P^{(G\beta)}(H) = \frac{1}{2^{N/2}}\left(\frac {\beta}{2\pi v^2}\right)^{\mu/2} 
                               \exp\left(-\frac{\beta}{4v^2}\Tr H^2\right)
\label{gauss}
\end{equation}
with the number $\mu$ of degrees of freedom given in
Eq.~(\ref{deg}). To properly account for the degeneracies in the
symplectic class, we define 
\begin{equation}
\Tr=\left\{\begin{array}{ll}\tr & \quad {\rm if} \quad \beta=1,2\\
                                           {\displaystyle\frac{1}{2}}\tr & \quad {\rm if} \quad \beta=4
               \end{array}\right. \ .
\label{trs}
\end{equation}
The probability density~(\ref{gauss}) and the symmetry class define
the Gaussian Orthogonal, Unitary and Symplectic Ensemble GOE, GUE and
GSE for $\beta=1,2,4$, respectively.  The generating function
\begin{eqnarray}
Z_k^{(G\beta)}(x+J,2v^2/\beta) &=& \int d[H] \, P^{(G\beta)}(H) 
                                                        \nonumber\\
 & & \qquad\qquad \prod_{p=1}^k \left( \frac{\det(H_0+\alpha H - x_p - J_p)}
                                              {\det(H_0+\alpha H - x_p + J_p)}\right)^\gamma 
\label{geng}
\end{eqnarray}
of this case has an exact representation as integral in superspace,
\begin{eqnarray}
Z_k^{(G\beta)}(x+J,2v^2/\beta) &=& \int d[\sigma] \, Q^{(G\beta)}(\sigma)   \nonumber\\
 & & \quad \detg^{-\beta/2\gamma}\left((\alpha\sigma-x-J)\otimes 1_{\gamma N} + 
                                                                                                1_{\zeta k}\otimes H_0\right) \ .
\label{gengs}
\end{eqnarray}
The crucial feature of supersymmetry is the drastic reduction in the
number of degrees of freedom. This is borne out in the dimension of
the matrix $\sigma$. It is a $2k\times 2k$ Hermitean supermatrix for
$\beta=2$ and a $4k\times 4k$ Hermitean supermatrix with additional
symmetries for $\beta=1,4$~\cite{EFE83}, we use the conventions of
Refs.~\cite{GUKOP2,GUKO2}. The parameter $\zeta=2$ for $\beta=2$ and
$\zeta=4$ for $\beta=1,4$ is defined accordingly. We write $1_M$ for
the $M\times M$ unit matrix. Thus, the expression~(\ref{gengs})
contains the unit matrices $1_N$ and $1_{4k}$ for $\beta=1$, $1_N$ and
$1_{2k}$ for $\beta=2$ and $1_{2N}$ and $1_{4k}$ for $\beta=4$.
Again, the volume element $d[\sigma]$ is the product of the
differentials of all independent variables. For the complex
anticommuting variables, we use the differentials of the variable and
of its complex conjugate. The probability density in superspace
\begin{equation}
Q^{(G\beta)}(\sigma) = c^{(\beta)} \exp\left(-\frac{\beta}{4v^2}\trg\sigma^2\right)
\label{supgauss}
\end{equation}
is a normalized Gaussian as well. Importantly, the normalization constants
\begin{equation}
c^{(\beta)}=\left\{\begin{array}{ll} 2^{k(k-1)}& \quad {\rm if} \quad \beta=2\\
                                                        2^{k(4k-3)/2}  & \quad {\rm if} \quad \beta=1,4
                            \end{array}\right. 
\label{noco}
\end{equation}
depend only on the dimension. In contrast to the ordinary
case~(\ref{gauss}), they doe not contain the variance $2v^2/\beta$.

The result~(\ref{gengs}) has a remarkable property. The
superdeterminant comprises a sum of two terms which are both direct
products. The first term is a direct product of supermatrices with a
unit matrix and ordinary space, and vice versa in the second term.
Most conveniently, this decouples to some extent the random matrix
ensemble, i.e.~the matrix $\sigma$ from the external field $H_0$.
This feature, which is typical for the supersymmetry method, was
already very helpful for an exact calculation of the transition from
Poisson regularity to the GUE in Refs.~\cite{Guh96,GUH4}.
Furthermore, it also made possible some asymptotic
evaluation~\cite{GuWe89,FGMG98} of the correlations on the local scale
for large coupling $\alpha/D$ where $D$ is the mean level spacing.

\subsection{Posing the Problem for Norm--dependent Ensembles}
\label{sec23} 

In analogy to the scalar product for vectors, one introduces a scalar
product $\Tr HK$ for two matrices $H$ and $K$ with the same symmetries.
This is then used to define the norm of a matrix by
\begin{equation}
\parallel{H}\parallel = \sqrt{\Tr H^2} \ ,
\label{norm}
\end{equation}
corresponding to the length of a vector. The class of norm--dependent
ensembles has a probability density of the form
\begin{equation}
P^{(\beta)}(H) = P^{(T\beta)}(\Tr H^2) \ ,
\label{pdnorm}
\end{equation}
where $P^{(T\beta)}(u)$ is function of the norm
$\parallel{H}\parallel$ or, equivalently of $u=\Tr H^2$. Of course,
$P^{(T\beta)}(u)$ has to be chosen such that $P^{(\beta)}(H)$ is
positive semi--definite and fulfills all the necessary convergence
requirements.  According to the symmetries, there are Norm--dependent
Orthogonal, Unitary and Symplectic Ensembles for $\beta=1,2,4$. We
denote them TOE, TUE and TSE, respectively. We show in~\ref{AppA} that
the $\nu$--th moment of the probability density can be expressed, if
it exists, as the single integral
\begin{eqnarray}
M_\nu^{(T\beta)} &=& \int P^{(T\beta)}(\Tr H^2) \left(\Tr H^2\right)^\nu d[H]
                                                        \nonumber\\
 &=&  \left(\frac{\pi}{2}\right)^{\mu/2} \frac{2^{N/2}}{\Gamma(\mu/2)}
          \int\limits_0^\infty u^{\nu+\mu/2-1} P^{(T\beta)}(u) du \ .
\label{mome}
\end{eqnarray}
This includes the normalization by setting $M_0^{(T\beta)}=1$ for
$\nu=0$.

Many ensembles fall into the norm--dependent class. Obviously, the
Gaussian Ensembles are found by setting
$P^{(T\beta)}(u)\sim\exp(-\beta u/4v^2)$, which is an exponential
function, not a Gaussian. Non--trivial examples are the fixed--trace
and the bound--trace ensembles~\cite{Mehta}. An important subclass of
norm--dependent ensembles is derived from a non--extensive entropy
principle~\cite{TVT04,BBP04}. It comprises a variety of interesting
cases which are found by considering limits of certain parameter. A
rather general construction of norm--dependent ensembles using a
single--valued spread function is given in Ref.~\cite{MK05}. We return
to this point.

We ask the following questions. Can we express the generating function
for the norm--dependent ensembles TOE, TUE and TSE given by
\begin{eqnarray}
Z_k^{(T\beta)}(x+J) &=& \int d[H] \, P^{(T\beta)}(\Tr H^2) 
                                                        \nonumber\\
 & & \qquad\qquad \prod_{p=1}^k \left( \frac{\det(H_0+\alpha H - x_p - J_p)}
                                              {\det(H_0+\alpha H - x_p + J_p)}\right)^\gamma 
\label{gent}
\end{eqnarray}
as integral in superspace? --- Can we construct the supersymmetric
analog of the probability density $P^{(T\beta)}(\Tr H^2) $ ? --- The
answers are in the affirmative. We will derive the exact
representation
\begin{eqnarray}
Z_k^{(T\beta)}(x+J) &=& \int d[\sigma]  \, Q^{(T\beta)}(\trg\sigma^2)   \nonumber\\
 & & \quad \detg^{-\beta/2\gamma}\left((\alpha\sigma-x-J)\otimes 1_{\gamma N} + 
                                                           1_{\zeta k}\otimes H_0\right) \ ,
\label{gents}
\end{eqnarray}
where the supermatrices $\sigma$ are defined as above and where the
probability density $Q^{(T\beta)}(\trg\sigma^2)$ is also
norm--dependent, but now in superspace.

Importantly, the direct product structure implying the decoupling of
the random matrix ensemble from the external field $H_0$ is also
present here for all TOE, TUE and TSE. This extends the discussion in
Section~\ref{sec22} for the Gaussian ensembles.

\section{Supersymmetric Representation}
\label{sec3} 

In Section~\ref{sec31}, we derive the supersymmetric representation by
using Fourier integrals. We present the transformation formulae for
the probability densities in Section~\ref{sec32}. The connection to
the spread function is discussed in Section~\ref{sec33}, which also
contains an alternative derivation of the transformation formulae.

\subsection{Derivation Using Fourier Integrals}
\label{sec31}

The norm $\parallel{H}\parallel$ is non--negative and we have $u=\Tr
H^2\ge 0$. Thus, $P^{(T\beta)}(u)$ is only defined on the positive $u$
axis.  When introducing the Fourier integral over the entire axis, we
have to set $P^{(T\beta)}(u)=0$ for $u<0$, such that
\begin{equation}
p^{(T\beta)}(y) = \frac{1}{\sqrt{2\pi}} 
                             \int\limits_0^\infty P^{(T\beta)}(u) \exp\left(iyu\right) du
\label{fou}
\end{equation}
is the Fourier transform with the inversion
\begin{equation}
P^{(T\beta)}(u) = \frac{1}{\sqrt{2\pi}} 
                             \int\limits_{-\infty}^{+\infty} p^{(T\beta)}(y) \exp\left(-iyu\right) dy \ .
\label{foui}
\end{equation}
We add a small imaginary increment to the Fourier variable,
$y^-=y-i\varepsilon$ and insert insert Eq.~(\ref{foui}) with $u=\Tr
H^2$, i.e.~the integral
\begin{equation}
P^{(T\beta)}(\Tr H^2) = \frac{1}{\sqrt{2\pi}} 
                             \int\limits_{-\infty}^{+\infty} p^{(T\beta)}(y) \exp\left(-iy^- \Tr H^2\right) dy \ .
\label{foug}
\end{equation}
into the generating function~(\ref{gent}).  We thereby rediscover the
Gaussian case~(\ref{geng}) with the variance $1/i2y^-$. The integrals
over $H$ can now be done as Gaussian integrals, the complex variance
$1/i2y^-$ does not cause a problem. Even without the imaginary
increment, they exist as Fresnel integrals. The imaginary increment
makes standard Gaussian integrals out of them, but this is not the
motivation for it. We need the imaginary increment later on. The
important difference to the Gaussian case of Section~\ref{sec22} is
the fact that the Gaussian $\exp\left(-iy^- \Tr H^2\right)$ comes
without normalization constant for the $H$ integration. Hence, when
inserting Eq.~(\ref{foug}) into Eq.~(\ref{geng}) we obtain the inverse
of the normalization constant as an $y$ dependent factor in the
Fourier integral,
\begin{eqnarray} 
Z_k^{(T\beta)}(x+J) &=& \frac{1}{\sqrt{2\pi}} 
                                          \int\limits_{-\infty}^{+\infty} dy \, p^{(T\beta)}(y) \nonumber\\ 
                                  & & \qquad\qquad 2^{N/2}\left(\frac{\pi}{i2y^-}\right)^{\mu/2}
                                         Z_k^{(G\beta)}(x+J,1/i2y^-) 
\label{gentf1}
\end{eqnarray}
with $Z_k^{(G\beta)}(x+J,1/i2y^-)$ given in Eq.~(\ref{geng}). We now
employ the supersymmetric representation~(\ref{gengs}) and find
\begin{eqnarray}
Z_k^{(T\beta)}(x+J) &=& \frac{1}{\sqrt{2\pi}} 
                                          \int\limits_{-\infty}^{+\infty} dy \, p^{(T\beta)}(y) 
                                         2^{N/2}\left(\frac{\pi}{i2y^-}\right)^{\mu/2}
                                                                          \nonumber\\
                                  & & \qquad \int d[\sigma]  \, c^{(\beta)}\exp\left(-iy^-\trg\sigma^2\right)
                                                                                                        \nonumber\\
                   & & \qquad\quad \detg^{-\beta/2\gamma}\left((\alpha\sigma-x-J)\otimes 1_{\gamma N} 
                           + 1_{\zeta k}H_0\right) \ .
\label{gentf2}
\end{eqnarray}
Hence, by interchanging the integrations, we arrive at the desired
Eq.~(\ref{gents}), where the probability density in superspace 
\begin{equation}
Q^{(T\beta)}(\trg\sigma^2) = c^{(\beta)} \frac{2^{N/2}}{\sqrt{2\pi}}
                                                 \left(\frac{\pi}{2}\right)^{\mu/2}
                             \int\limits_{-\infty}^{+\infty} p^{(T\beta)}(y) 
                             \frac{\exp\left(-iy^- \trg\sigma^2\right)}{(iy^-)^{\mu/2}} dy \ .
\label{probs}
\end{equation}
is the inverse Fourier integral with an additional power
$(iy^-)^{\mu/2}$ in the denominator.

\subsection{Transformation Formulae}
\label{sec32}

We set $w=\trg\sigma^2$ and plug the Fourier integral~(\ref{fou}) into
Eq.~(\ref{probs}),
\begin{equation}
Q^{(T\beta)}(w) = c^{(\beta)} \frac{2^{N/2}}{2\pi}
                                                 \left(\frac{\pi}{2}\right)^{\mu/2}
                             \int\limits_0^{\infty} du \, P^{(T\beta)}(u) 
                             \int\limits_{-\infty}^{+\infty}
                             \frac{\exp\left(-iy(u-w)\right)}{(iy^-)^{\mu/2}} dy \ .
\label{probsf}
\end{equation}
The $y$ integral converges because of the imaginary increment and can
be done in a standard way~\cite{GR}. Apart from factors, it yields
$\Theta(u-w)(u-w)^{\mu/2-1}$. We thus arrive at the transformation formula
\begin{equation}
Q^{(T\beta)}(w) = \frac{c^{(\beta)}2^{N/2}}{\Gamma(\mu/2)}
                                                 \left(\frac{\pi}{2}\right)^{\mu/2}
                             \int\limits_0^{\infty} P^{(T\beta)}(u+w) u^{\mu/2-1} du \ .
\label{trafos}
\end{equation}
This result allows one to calculate, by a single integration, the
probability density in superspace for any norm--dependent ensemble
TOE, TUE and TSE. We notice that the Fourier integral~(\ref{fou}) has
to exist, but, importantly, its explicit knowledge is not needed to
obtain the probability density $Q^{(T\beta)}(w)$ in superspace.
Interestingly, the transformation formula can be inverted. For even
number of degrees of freedom $\mu$, iterated integration by parts
yields
\begin{equation}
P^{(T\beta)}(u) = \frac{(-1)^{\mu/2}}{c^{(\beta)}2^{N/2}}
                                                 \left(\frac{2}{\pi}\right)^{\mu/2}
                             \frac{\partial^{\mu/2}}{\partial u^{\mu/2}} Q^{(T\beta)}(u) \ .
\label{trafoo}
\end{equation}
This inversion is likely to be correct even for odd $\mu$ if the
theory of fractional derivatives is applied. 

From a conceptual viewpoint, the pair of transformation
formulae~(\ref{trafos}) and~(\ref{trafoo}) states the main result of
this contribution. The power of supersymmetry lies in the drastic
reduction of the degrees of freedom. The mechanism of how this happens
was previously only known in the Gaussian case. The transformation
formulae~(\ref{trafos}) and~(\ref{trafoo}) considerably generalize
that.  A particularly interesting interpretation follows from
formula~(\ref{trafoo}).  The probability densities $P^{(T\beta)}(u)$
and $Q^{(T\beta)}(u)$ formally coincide for $\mu=0$, i.e.~in zero
dimensions, $N=0$. This is already visible in the Gaussian case.
Apart from the variance independent normalization $c^{(\beta)}$, the
Gaussian~(\ref{supgauss}) in superspace with $u=\trg\sigma^2$ indeed
follows from the Gaussian~(\ref{gauss}) in ordinary space with $u=\Tr
H^2$ by simply setting $N=0$.

Another interesting observation results from putting $w=0$ in the
transformation formula~(\ref{trafos}) and then using the normalization
of $P^{(T\beta)}(u)$ which can be read off from Eq.~(\ref{mome}) for
$\nu=0$. One has $Q^{(T\beta)}(0)=c^{(\beta)}$. In other words, the
normalization of the probability density in ordinary space corresponds
to the value of the probability density in superspace at $w=0$. This
reflects the Efetov--Wegner--Parisi--Sourlas
theorem~\cite{EFE83,PaSou,Weg,ConGro}, referred to as Rothstein
theorem in mathematics~\cite{ROT}. It implies that the normalization
integral for a function such as ours which only depends on invariants
reads
\begin{equation}
1 = \int Q^{(T\beta)}(\trg\sigma^2) d[\sigma] = \frac{1}{c^{(\beta)}} Q^{(T\beta)}(0) \ .
\label{efewe}
\end{equation}
This phenomenon exclusively occurs in superspace is due to a subtle
mutual cancellation of singularities. Hence, it is reassuring to see
that the normalization of the probability density in ordinary space
leads --- via the Efetov--Wegner--Parisi--Sourlas theorem --- to the
normalization of the probability density in superspace.

\subsection{Connection to the Spread Function}
\label{sec33}

A rather general construction of norm--dependent ensembles was given
by Muttalib and Klauder~\cite{MK05} for the unitary case. It can be
generalized to all three symmetry classes in a straightforward
manner. The probability density
\begin{equation}
P^{(T\beta)}(\Tr H^2) = \int\limits_0^\infty f^{(T\beta)}(t) \,
                               \frac{1}{2^{N/2}}\left(\frac {\beta}{2\pi t}\right)^{\mu/2} 
                               \exp\left(-\frac{\beta}{4t}\Tr H^2\right) \, dt
\label{spread}
\end{equation}
is expressed as an integral involving a normalized Gaussian with a
real variance $2t/\beta$. The quantity $f^{(T\beta)}(t)$ is referred to
as spread function. As seen form Eq.~(\ref{spread}) it is normalized,
\begin{equation}
\int\limits_0^\infty f^{(T\beta)}(t) \, dt = 1 \ .
\label{spreadnorm}
\end{equation}
We insert the integral~(\ref{spread}) into the generating
function~(\ref{gent}) and find in steps analogous to the ones in
Section~\ref{sec31},
\begin{eqnarray}
Z_k^{(T\beta)}(x+J) &=& \int\limits_0^\infty dt \, f^{(T\beta)}(t) \,
                                        Z_k^{(G\beta)}(x+J,2t/\beta)
                                                                          \nonumber\\
                                  &=& \int\limits_0^\infty dt \, f^{(T\beta)}(t)  
                                           \int d[\sigma]  \, c^{(\beta)}\exp\left(-\frac{\beta}{4t}\trg\sigma^2\right)
                                                                                                        \nonumber\\
                   & & \qquad \detg^{-\beta/2\gamma}\left((\alpha\sigma-x-J)\otimes 1_{\gamma N} 
                           + 1_{\zeta k}H_0\right) \ .
\label{gensp}
\end{eqnarray}
This yields again Eq.~(\ref{gents}) where the probability density in superspace now
reads
\begin{equation}
Q^{(T\beta)}(\trg\sigma^2) = \int\limits_0^\infty f^{(T\beta)}(t) \, c^{(\beta)}
                                                \exp\left(-\frac{\beta}{4t}\trg\sigma^2\right) \, dt \ .
\label{spreads}
\end{equation}
Comparing Eqs.~(\ref{spread}) and~(\ref{spreads}) one sees that the
probability densities are in ordinary and in superspace given as
integrals over the spread function times a normalized
Gaussian. Moreover, we notice that the variable $t$ in the variance
$2t/\beta$ has the meaning of a diffusion time. In ordinary space, the
diffusion is Dyson's Brownian Motion~\cite{DYS1,DYS2}. It has a fully
fledged analog in superspace~\cite{GUH4} with the same diffusion time.
Thus, the TOE, TUE and TSE are, in ordinary and in superspace,
ensembles constructed as averages involving the diffusion time.

The transformation formulae~(\ref{trafos}) and~(\ref{trafoo}) are
easily re--derived from Eqs.~(\ref{spread}) and~(\ref{spreads}).  We
emphasize that only the existence, but not the precise knowledge of
the spread function is needed to calculate the probability density in
superspace. Those readers might appreciate the alternative derivation
of the transformation formulae by means of the spread function who did
not feel comfortable with our treatment of the singularities in the
Fourier integrals of Section~\ref{sec31}.

\section{Some Specific Examples}
\label{sec4} 

To gain insight into how the transformation formulae work, we
calculate the probability densities in superspace for a variety of
examples.  To acquire some first experience, we revisit the Gaussian
ensembles in Section~\ref{sec41}. We discuss, for all three symmetry
classes $\beta=1,2,4$ the bound trace, the fixed trace, the
Gauss--monomial and the Gauss--quartic ensembles in
Sections~\ref{sec42} to~\ref{sec45}. For the probability densities in
ordinary space of these examples, we introduce constants $a_0$, $a_1$
and $a_2$ which are always assumed to be real and positive. Using
Eq.~(\ref{mome}), they can be expressed in terms of the moments
$M_\nu^{(T\beta)}$. In particular, the overall normalization constant
can be fixed with Eq.~(\ref{mome}) for $\nu=0$.  However, we rather
use the relation $Q^{(T\beta)}(0)=c^{(\beta)}$ which is according to
Section~\ref{sec32} equivalent to the normalization of the probability
density in ordinary space. In Section~\ref{sec46} we discuss the
ensembles derived from an non--extensive entropy principle. We always
write $u=\Tr H^2$ and $w=\trg\sigma^2$.

\subsection{Revisiting the Gaussian Ensembles}
\label{sec41}

Inserting the Gaussian~(\ref{gauss}) into the transformation
formula~(\ref{trafos}), we find
\begin{eqnarray}
Q^{(G\beta)}(w) &=& \exp\left(-\frac{\beta}{4v^2}w\right)
                                    \frac{c^{(\beta)}}{\Gamma(\mu/2)}
                                    \left(\frac {\beta}{4v^2}\right)^{\mu/2} 
                                                              \nonumber\\
                            & & \qquad\qquad
                                    \int\limits_0^{\infty} \exp\left(-\frac{\beta}{4v^2}u\right) u^{\mu/2-1} du \ ,
\label{trafogauss}
\end{eqnarray}
which gives the Gaussian~(\ref{supgauss}). The Fourier transform
\begin{equation}
p^{(G\beta)}(y) =  \frac{1}{\sqrt{2\pi}2^{N/2}}
                              \left(\frac{\beta}{2\pi v^2}\right)^{\mu/2} 
                              \frac{1}{iy-\beta/4v^2} 
\label{fougauss}
\end{equation}
can be also be used to infer the spread function, which is a $\delta$
function. The integral~(\ref{spread}) has to be interpreted as a
proper Cauchy integral.

\subsection{Bound Trace Ensembles}
\label{sec42}

The probability density sets a cutoff for the norm of the random matrices
according to~\cite{Mehta}
\begin{equation}
P^{(BT\beta)}(u) = a_0 \Theta\left(a_1-u\right) \ .  
\label{bnd}
\end{equation}
The transformation formula~(\ref{trafos}) yields
\begin{equation}
Q^{(BT\beta)}(w) = c^{(\beta)} \frac{(a_1-w)^{\mu/2}}{a_1^{\mu/2}} 
                                                                          \Theta\left(a_1-w\right) \ ,
\label{bnds}
\end{equation}
which is a bound trace ensemble as well, but now in superspace and
multiplied with a polynomial factor.

\subsection{Fixed Trace Ensembles}
\label{sec43}

The probability density fixes the norm of the random matrices such
that~\cite{Mehta,LCD99,ACMV99}
\begin{equation}
P^{(FT\beta)}(u) = a_0 \delta\left(a_1-u\right) \ .  
\label{fix}
\end{equation}
With the transformation formula~(\ref{trafos}) we find
\begin{equation}
Q^{(FT\beta)}(w) = c^{(\beta)} \frac{(a_1-w)^{\mu/2-1}}{a_1^{\mu/2-1}} 
                                                                          \Theta\left(a_1-w\right) \ ,
\label{fixs}
\end{equation}
which is, once more, a bound trace ensemble of the form~(\ref{bnds}).
We notice that the exponent in the polynomial factor is $\mu/2-1$
compared with $\mu/2$ in Eq.~(\ref{bnds}). This simply reflects that
the probability density~(\ref{fix}) is the derivative of the
probability density~(\ref{bnd}). We mention that fixed trace ensembles
do not seem to exist in superspace, at least not in a simple--minded
interpretation. This is so, because the normalization requirement
$Q^{(T\beta)}(0)=c^{(\beta)}$ can hardly be fulfilled if
$Q^{(T\beta)}(w)$ includes a $\delta$ function.

\subsection{Gauss--Monomial Ensembles}
\label{sec44}

The probability densities in superspace derived in the previous
examples tend to have remarkable similarity to the ones in ordinary
space. This seems to be a fairly robust phenomenon. To illustrate it
further, we introduce ensembles comprising a Gaussian and a monomial
factor,
\begin{equation}
P^{(GM\beta)}(u) = a_0 u^m \exp\left(-a_1u\right) \ ,  
\label{gm}
\end{equation}
where $m$ is an integer.  The transformation formula~(\ref{trafos})
implies
\begin{equation}
Q^{(GM\beta)}(w) = c^{(\beta)} \exp\left(-a_1w\right) 
                               \sum_{m'=0}^m \left(\begin{array}{c} m \\
                                                                                             m'
                                                          \end{array}\right) 
                               \frac{\Gamma(m-m'+\mu/2)}{\Gamma(m+\mu/2)} (a_1w)^{m'} \ .
\label{gms}
\end{equation}
These are Gauss--polynomial ensembles including all powers between
zero and $m$.

\subsection{Gauss--Quartic Ensembles}
\label{sec45}

We now consider Gaussian probability densities supplemented with a
quartic term in the exponent,
\begin{equation}
P^{(GQ\beta)}(u) = a_0 \exp\left(-a_1u -a_2u^2\right) \ .  
\label{gq}
\end{equation}
With the transformation formula~(\ref{trafos}), we obtain
\begin{equation}
Q^{(GQ\beta)}(w) = c^{(\beta)} \exp\left(-\frac{a_1}{2}w-\frac{a_2}{2}w^2\right) 
                               \frac{D_{-\mu/2}\left(a_1/\sqrt{2a_2}+\sqrt{2a_2}w\right)}
                                      {D_{-\mu/2}\left(a_1/\sqrt{2a_2}\right)} \ ,
\label{gqs}
\end{equation}
where $D_p(z)$ denotes the parabolic cylinder function of order
$p$~\cite{GR}. Once more, the probability density in superspace
contains the functional form of the one in ordinary space. However,
this example also shows that one can come up with cases in which the
additional contribution has a rather inconvenient structure.

\subsection{Ensembles Deriving from a Non--extensive Entropy Principle}
\label{sec46}

An interesting family of ensembles was constructed in
Refs.~\cite{TVT04,BBP04}.  Among other features, it yields in a
certain parameter range an invariant L{\'e}vy--type--of ensemble.  The
construction of Refs.~\cite{TVT04,BBP04} is done for the orthogonal
symmetry class, but can easily be generalized to all $\beta=1,2,4$.
The probability density in ordinary space
\begin{equation}
P^{(NE\beta)}(u) = a_0 \left(1+\frac{\kappa}{\Lambda}u\right)^{1/(1-q)}   
\label{ne}
\end{equation}
depends on a parameter $q$ used in the non--extensive entropy.
Moreover, it includes a positive parameter $\kappa$ and
\begin{equation}
\Lambda = \frac{1}{q-1} - \frac{\mu}{2} \ ,
\label{nep}
\end{equation}
with $\mu$ being the number of degrees of freedom~(\ref{deg}). To
avoid confusion with the notation in the present contribution, we
write $\kappa$, $\Lambda$, $\mu$, instead of $\alpha$, $\lambda$, $f$
in Ref.~\cite{BBP04}.  

We consider $q>1$. This choice makes the exponent in the probability
density~(\ref{ne}) negative. Moreover, it requires $\Lambda>0$ such
that
\begin{equation}
1 < q < q_{{\rm max}} = 1 + \frac{2}{\mu} \ .
\label{neq}
\end{equation}
With help of the integral representation~\cite{BBP04}
\begin{equation}
P^{(NE\beta)}(u) = \frac{a_0}{\Gamma(1/(q-1))} \int\limits_0^\infty \xi^{1/(q-1)-1}
                                \exp\left(-\left(1+\frac{\kappa}{\Lambda}u\right)\xi\right) d\xi \ ,
\label{nei}
\end{equation}
we can obtain the probability density in superspace from the
transformation formula~(\ref{trafos}). The $u$ integral has to be done
first. Convergence is ensured because of the condition $\Lambda>0$.
Collecting everything, we arrive at
\begin{equation}
Q^{(NE\beta)}(w) = c^{(\beta)} \left(1+\frac{\kappa}{\Lambda}w\right)^{-\Lambda} \ . 
\label{nes}
\end{equation}
Remarkably, this is again the same functional form as in ordinary
space. The (negative) exponent $1/(1-q)$ in ordinary space is mapped
onto $-\Lambda$ in superspace. We notice that $Q^{(NE\beta)}(w)$
depends on $q$ only via the parameter $\Lambda$ which appears twice in
Eq.~(\ref{nes}).

\section{Correlation Functions}
\label{sec5} 

After discussing general results for all symmetry classes in
Section~\ref{sec51}, we give more explicit formulae for the unitary
case in Section~\ref{sec52}.  All results to be given here can also be
averaged over the external field $H_0$ with some probability density
$P_0(H_0)$. However, as this is an obvious step, we do not go into
that further.

\subsection{All Symmetry Classes}
\label{sec51} 

We now have the supersymmetric representation~(\ref{gents}) for the
generating function and the one--dimensional transformation
formula~(\ref{trafos}) for the probability density $Q^{(T\beta)}(w)$
in superspace. Hence, we can exploit the advantages of supersymmetry
also for norm--dependent ensembles. In particular, the level number
$N$ is, in contrast to the ordinary space, an explicit parameter in
Eq.~(\ref{gents}). Depending on the ensemble, this can makes it
possible to study the generating function by means of a saddle point
approximation for large $N$. As the details of such a calculation will
sensitively depend on the specific form of $Q^{(T\beta)}(w)$, we
refrain from attempting a general discussion.

We can also proceed by observing that Eqs.~(\ref{gentf1})
and~(\ref{gensp}) are integral transforms of the generating functions
involving the Fourier integral,
\begin{equation} 
Z_k^{(T\beta)}(x+J) = \frac{2^{N/2}\pi^{\mu/2}}{\sqrt{2\pi}} 
                                          \int\limits_{-\infty}^{+\infty} \frac{p^{(T\beta)}(y)}{(i2y^-)^{\mu/2}} \,
                                          Z_k^{(G\beta)}(x+J,1/i2y^-) \, dy \ ,
\label{ztrafo1}
\end{equation}
or the spread function,
\begin{equation} 
Z_k^{(T\beta)}(x+J) = \int\limits_0^\infty f^{(T\beta)}(t) \, Z_k^{(G\beta)}(x+J,2t/\beta) \, dt \ ,
\label{ztrafo2}
\end{equation}
respectively. Thus, the correlation functions
$R_k^{(T\beta)}(x_1,\ldots,x_k,\alpha,H_0)$ of all norm--dependent
ensembles are obtained as single integrals over the corresponding ones
$R_k^{(G\beta)}(x_1,\ldots,x_k,\alpha,H_0)$ in the Gaussian case. With
Eqs.~(\ref{corr}) and ~(\ref{ztrafo1}), we find
\begin{eqnarray}
&&R_k^{(T\beta)}(x_1,\ldots,x_k,\alpha,H_0)           
                                                    \nonumber\\
& & \qquad = \frac{2^{N/2}\pi^{\mu/2}}{\sqrt{2\pi}} \int\limits_{-\infty}^{+\infty} dy \,
                                          \frac{p^{(T\beta)}(y)}{(i2y^-)^{\mu/2}}\left(\frac{i2y}{\beta}\right)^{k/2}
                                                    \nonumber\\
& & \qquad\qquad\qquad
          R_k^{(G\beta)}(x_1\sqrt{2iy/\beta},\ldots,x_k\sqrt{2iy/\beta},\alpha,H_0\sqrt{2iy/\beta}) \  .
\label{corrf}
\end{eqnarray}
In those cases where the imaginary unit and the singularities can cause problems, one
should rather resort to the alternative expression deriving from Eq.~(\ref{ztrafo2}),
\begin{eqnarray}
&&R_k^{(T\beta)}(x_1,\ldots,x_k,\alpha,H_0) 
                                                                         \nonumber\\
& & \qquad = \int\limits_0^\infty dt \, \frac{f^{(T\beta)}(t)}{(2t)^{k/2}}
      R_k^{(G\beta)}(x_1/\sqrt{2t},\ldots,x_k/\sqrt{2t},\alpha,H_0/\sqrt{2t}) \  .
\label{corrs}
\end{eqnarray}
For $H_0=0$ and $\beta=1$, Eq.~(\ref{corrs}) was already obtained in
Ref.~\cite{MK05} and a similar result is given in Ref.~\cite{BBP04}
for the ensembles deriving from non--extensive entropy. Here, this is
generalized to all three symmetry classes.

Of course, supersymmetry is not needed to derive
formulae~(\ref{corrf}) or~(\ref{corrs}). For $H_0=0$, one can now use
the closed expressions for the correlation functions
$R_k^{(G\beta)}(x_1,\ldots,x_k,1,0)$ from Ref.~\cite{Mehta} and
calculate the correlation functions for all ensembles TOE, TUE and
TSE. For $H_0\neq 0$, however, supersymmetry is very helpful, because
it provides formula~(\ref{gengs}). As already discussed in
Section~\ref{sec22}, the random matrix ensemble, i.e.~the matrix
$\sigma$, is to some extent decoupled from the external field $H_0$
due to the direct product structure.  This makes it possible to obtain
asymptotic results for large coupling $\alpha/D$ as in
Refs.~\cite{GuWe89,FGMG98} which can then be inserted into
formulae~(\ref{corrf}) and~(\ref{corrs}).

\subsection{Unitary Symmetry Class}
\label{sec52} 

In the unitary symmetry class, i.e.~for $\beta=2$, we can gain
additional insights by following the steps outlined in
Refs.~\cite{TG,Guh96,GUH4}.  We work with the integral
transform~(\ref{ztrafo2}) and the supersymmetric representation of the
generating function~(\ref{gengs}). We absorb the parameter $\alpha$
into the supermatrix $\sigma$, which is equivalent to multiplying the
variance of the Gaussian probability density with $\alpha^2$.  The
supermatrix $\sigma$ is now shifted by $x+J$ to remove these latter
matrices from the superdeterminant. We then diagonalize
$\sigma=usu^{-1}$ and do the angular integration over the unitary
supermatrix $u$. Here, $s=\diag(s_{11},is_{12},\ldots,s_{k1},is_{k2})$
is a diagonal matrix containing the eigenvalues $s_{p1} , \
p=1,\ldots,k$ in the bosonic and $is_{p2} , \ p=1,\ldots,k$ in the
fermionic sector.  This yields~\cite{Guh96,GUH4}
\begin{eqnarray}
Z_k^{(G2)}(x+J,t) &=& 1 - \eta(x+J)  
\nonumber\\
 & & \quad+ \frac{1}{B_k(x+J)} \int d[s] \, B_k(s)  
                                                   \nonumber\\
 & & \quad\quad \frac{1}{(2\pi t\alpha^2)^k}\exp\left(-\frac{1}{2t\alpha^2}\trg(s-x-J)^2\right)
                                                   \nonumber\\
 & & \quad\quad\quad \detg^{-1}\left(s^\pm \otimes 1_N - 1_{2k}\otimes H_0\right) \ ,
\label{gengse}
\end{eqnarray}
where the function
\begin{equation}
B_k(s) = \det\left[\frac{1}{s_{p1}-is_{q2}}\right]_{p,q=1,\ldots,k} 
\label{bks}
\end{equation}
is the square root of the Jacobian or Berezinian arising when changing
the integration variables in superspace to eigenvalue--angle
coordinates. The eigenvalues in the supermatrix carry a small
imaginary increment to ensure convergence, we write $s^\pm$. The
function $\eta(x+J)$ in Eq.~(\ref{gengse}) takes care of some
Efetov--Wegner--Parisi--Sourlas contributions which are not needed
here as we are mainly interested in the correlations of the imaginary
parts.

Inserting Eq.~(\ref{gengse}) into the integral transform~(\ref{ztrafo2}) and using the
normalization~(\ref{spreadnorm}), we arrive at
\begin{eqnarray}
Z_k^{(T2)}(x+J) &=& 1 - \eta(x+J)  
\nonumber\\
 & & \quad + \frac{1}{B_k(x+J)} \int d[s] \, B_k(s)    
                                                   \nonumber\\
 & & \quad\quad \frac{1}{\alpha^{2k}}Q_E^{(T2)}\left(\frac{1}{\alpha^2}\trg(s-x-J)^2\right)
                                                   \nonumber\\
 & & \quad\quad\quad \detg^{-1}\left(s^\pm \otimes 1_N - 1_{2k}\otimes H_0\right) \ ,
\label{gengteq}
\end{eqnarray}
where we introduced the probability density 
\begin{equation}
Q_E^{(T2)}(w) = \int\limits_0^\infty f^{(T2)}(t) \, \frac{1}{(2\pi t)^k} \, 
                                              \exp\left(-\frac{w}{2t}\right) \, dt \ .
\label{spreadse}
\end{equation}
Both of the expressions~(\ref{spreads}) and~(\ref{spreadse}) yield
probability densities in superspace, the former in the full, the
latter in the eigenvalue space. This is the reason why Eq.~(\ref{spreadse}) contains
the power $t^k$ in the denominator, it arose from the integration over the
unitary supermatrix. Nevertheless, we can apply the same line of arguing as
in Section~\ref{sec33} and derive from Eqs.~(\ref{spread}) and~(\ref{spreadse})
the transformation formula 
\begin{equation}
Q_E^{(T2)}(w) = \frac{2^{N/2}\pi^{(N^2-2k)/2}}{\Gamma((N^2-2k)/2)}
                             \int\limits_0^{\infty} P^{(T2)}(u+w) u^{(N^2-2k)/2-1} du \ ,
\label{trafose}
\end{equation}
with $\mu=N^2$ being the number of degrees of freedom in the unitary
case $\beta=2$. A comparison with the transformation
formula~(\ref{trafos}) shows that the probability densities in the
eigenvalue superspace follows from the one in full superspace by
simply replacing the number of degrees of freedom $N^2$ with the
reduced number $N^2-2k$ where $2k$ is the number of degrees of freedom
in the eigenvalue superspace,
\begin{equation}
Q_E^{(T2)}(w) = \frac{2^{(N^2-2k)/2}}{c^{(2)}} \, Q^{(T2)}(w) \Bigg|_{\mu=N^2-2k} \ .
\label{trafoser}
\end{equation}
One also obtains the inversion of Eq.~(\ref{trafose}) by modifying the transformation
formula~(\ref{trafoo}) accordingly.

Hence, we now have an exact expression~(\ref{gengteq}) of the
generating function for all TUE as a $2k$ dimensional integral with a
probability density given by Eqs.~(\ref{trafose}) or~(\ref{trafoser}).
Before going over to the correlation functions, a caveat of the same
kind as discussed in Ref.~\cite{GUH4} is in order.  The limit
$\alpha\to 0$ can be taken without problems in Eq.~(\ref{gengteq}),
thereafter the derivatives with respect to the source variables $J$
can be taken and yield the correlation functions for
$\alpha=0$. Because of some interference with the
Efetov--Wegner--Parisi--Sourlas term, this should not be done in
reversed order for the case $\alpha=0$. Thus, the following discussion
applies to $\alpha>0$, where the derivatives of the generating
function~(\ref{gengteq}) can be taken first. If we may assume that the
probability density~(\ref{trafose}) does not contain inverses of
source variables, we find as in Refs.~\cite{TG,Guh96,GUH4}
\begin{eqnarray}
R_k^{(T2)}(x_1,\ldots,x_k,\alpha,H_0) &=& \frac{(-1)^k}{\pi^k} \int d[s] \, B_k(s)    
                \, \frac{1}{\alpha^{2k}}Q_E^{(T2)}\left(\frac{1}{\alpha^2}\trg(s-x)^2\right)
                                                   \nonumber\\
 & & \qquad\qquad\qquad \Im\prod_{p=1}^k \prod_{n=1}^N
                                     \frac{is_{p2} - H_{0n}}{s_{p1}^\pm  - H_{0n}}  \ ,
\label{gengtcorr}
\end{eqnarray}
where $H_{0n}$ are the entries of the diagonal matrix $H_0$. The
symbol $\Im$ denotes the proper restriction to the imaginary parts
which will be explained below. As already observed in Ref.~\cite{TG},
the determinantal structure of the GUE correlation functions arises in
this supersymmetry approach as an immediate consequence of the
determinant structure~(\ref{bks}). Thus, it follows form the Jacobian
in superspace. The only term in the integrand which can destroy this
feature for the TUE is the probability density $Q_E^{(T2)}(w)$.  To
circumvent this problem, we use formula~(\ref{spreadse}) which makes
possible to advantage of explicit results for the GUE correlation
functions in the presence of an external field.  The formulae given in
Ref.~\cite{Guh96,GUH4} are slightly more general than what we need in
the present context, because they also contains an integral over the
probability density $P_0(H_0)$ of the external field $H_0$. However, a
$\delta$ function $P_0(H_0)$ trivially yields
\begin{equation}
R_k^{(G2)}(x_1,\ldots,x_k,t\alpha^2,H_0) = 
\det\left[C_N(x_p,x_q,t\alpha^2,H_0)\right]_{p,q=1,\ldots,k} \ .
\label{guecorr}
\end{equation}
The kernel is given as the double integral 
\begin{eqnarray}
&&C_N(x_p,x_q,t\alpha^2,H_0) 
                                                                         \nonumber\\
&& \qquad = -\frac{1}{2\pi^2 t\alpha^2} \int\limits_{-\infty}^{+\infty}\int\limits_{-\infty}^{+\infty}
                      \frac{ds_1ds_2}{s_1-is_2}\exp\left(\frac{(is_2-x_q)^2}{2t\alpha^2}
                                                                                    -\frac{(s_1-x_p)^2}{2t\alpha^2}\right)
                                                                        \nonumber\\
&& \qquad\qquad\qquad\qquad\qquad\qquad
                                     \Im \prod_{n=1}^N \frac{is_2-H_{0n}}{s_1^--H_{0n}}  \ ,        
\label{gueker}
\end{eqnarray}
where we drop the indices $p$ and $q$ of the integration
variables. The present notation slightly deviates, in a hopefully
self--explanatory way, from the previous one. We now have the variance
$t\alpha^2$ as an argument, because it contains the parameter $\alpha$
after the above mentioned changes of integration variables.  The
process of going over to the imaginary parts of the correlation
functions amounts to inserting the definition
\begin{equation}
\Im \prod_{n=1}^N \frac{is_2-H_{0n}}{s_1^--H_{0n}}  = \frac{1}{i2} \left(
\prod_{n=1}^N \frac{is_2-H_{0n}}{s_1^--H_{0n}} -
                      \prod_{n=1}^N \frac{is_2-H_{0n}}{s_1^+-H_{0n}} \right) \ .
\label{ima}
\end{equation}
We notice that the term $1/(s_1-is_2)$ in the integrand of
Eq.~(\ref{gueker}) is the remainder of the Jacobian. After
Refs.~\cite{TG,Guh96,GUH4}, such double integral expressions were also
derived by other authors.

We combine these findings and arrive at
\begin{eqnarray}
& & R_k^{(T2)}(x_1,\ldots,x_k,\alpha,H_0) 
                                          \nonumber\\ 
& & \qquad =  \int\limits_0^\infty dt \, \frac{f^{(T2)}(t)}{(2\pi t)^k} 
                        \det\left[C_N(x_p,x_q,t\alpha^2,H_0)\right]_{p,q=1,\ldots,k} 
\label{gengtcorrf}
\end{eqnarray}
which is an exact representation of the TUE correlation functions for
finite level number in the presence of an external field. For
convenience, we derived this result using the spread function
$f^{(T2)}(t)$.  However, depending on the specific form of the
probability density $Q_E^{(T2)}(w)$ given in Eqs.~(\ref{trafose})
and~(\ref{trafoser}), one might want to prefer another integral
representation. Any such representation which involves a Gaussian will
lead to a result of the form~(\ref{gengtcorrf}), but with the spread
function replaced by another function. It should also be mentioned
that the result~(\ref{gengtcorrf}) can alternatively be derived
starting directly from Eq.~(\ref{corrs}). This, however, would not
lead to the probability density $Q_E^{(T2)}(w)$ and the
corresponding observation concerning the reduced number of degrees of
freedom. Furthermore, we notice that the double
integral~(\ref{gueker}) for the kernel can be evaluated in closed
form. For the sake of completeness, we give this result in~\ref{AppB}.

\section{Summary and Conclusions}
\label{sec6}

We showed that the all norm--dependent random matrix ensembles TOE,
TUE and TSE have a supersymmetric representation. Hence, supersymmetry
is by no means restricted to Gaussian probability densities. We mapped
the functions generating the $k$ point correlation functions onto
their supersymmetric analogues.  All our results include an external
field.  No approximation was made, all expressions are exact and for
finite level number. We derived transformation formulae which yield
the probability density in superspace as one--dimensional integral
involving the probability density in ordinary space.  These formulae
state the most important conceptual insight of this contribution.  We
emphasize that only the existence, but not the explicit knowledge of
the Fourier integral or the spread function is needed to obtain the
probability density in superspace.  The transformation formulae
clarify the mechanism of how supersymmetry manages to reduce the
number of degrees of freedom.  We worked out several
examples. Remarkably, the functional forms of the probability
densities tends to be very similar in ordinary and superspace. In
particular, this is so for the whole family of ensembles deriving from
a non--extensive entropy principle.

{}From a practical viewpoint, our most important findings are a series
of exact results for the correlation functions which can be used in
applications.  Different limits for the level number or other
parameters can now be studied depending on the ensemble of
interest. The most explicit formulae are for the TUE where we employed
the determinant structure of the GUE correlations. We also derived a
probability density for the TUE in the superspace of eigenvalues.

Can supersymmetry be applied to ensembles which are more general than
the norm--dependent ones? --- Indeed, a supersymmetric representation
is possible under quite general conditions. However, as this
construction requires a completely different approach, we defer it to
another contribution~\cite{TGhst}.

\section*{Acknowledgments}

I thank Frieder Kalisch and Heiner Kohler for fruitful discussions.  I
acknowledge financial support from Det Svenska Vetenskapsr\aa det.

\appendix

\section{Moments of Norm--dependent Probability Densities}
\label{AppA}

The following calculation, although rather straightforward, yields an
interesting side result which might be useful in other
applications. This is why we sketch the calculation here. Inserting
the diagonalizations $H=UxU^{-1}$ with $x=\diag(x_1,\ldots,x_N)$ and a
double degeneracy for $\beta=4$, we find from the definition of the
moments in Eqs.~(\ref{mome})
\begin{eqnarray}
M_\nu^{(T\beta)} &=& \pi^{\beta N(N-1)/4}
                              \frac{\Gamma^N(1+\beta/2)}{\prod_{n=1}^N\Gamma(1+n\beta/2)}
                                                                  \nonumber\\
       & & \qquad\qquad
                              \int P^{(T\beta)}(\Tr x^2) \left(\Tr x^2\right)^\nu |\Delta_N(x)|^\beta d[x] \ ,
\label{a1}
\end{eqnarray}
where $\Delta_N(x)=\prod_{n<m}(x_n-x_m)$ is the Vandermonde
determinant. The constant in front of the eigenvalue integral contains
the result of the integration over $U$ and also some factors stemming
from the Jacobian of the transformation to eigenvalue--angle
coordinates. We view the eigenvalues as components of a vector
$\vec{r}=(x_1,\ldots,x_N)$ in $N$ dimensions and introduce polar
coordinates $\vec{r}=r\vec{e}$ where $r$ is the length and $\vec{e}$ a
unit vector depending on $N-1$ angles.  The volume element reads
$d[x]=d^Nx=r^{N-1} dr d\Omega$ where $d\Omega$ is the infinitesimal
solid angle.  Because of $r^2=\Tr x^2$, we have
\begin{eqnarray}
M_\nu^{(T\beta)} &=& \pi^{\beta N(N-1)/4}
                              \frac{\Gamma^N(1+\beta/2)}{\prod_{n=1}^N\Gamma(1+n\beta/2)}
                                                                  \nonumber\\
       & & \qquad\qquad
                              \int |\Delta_N(\vec{e})|^\beta d\Omega
                              \int\limits_0^\infty r^{\nu+N-1+\beta N(N-1)/2} P^{(T\beta)}(r^2) dr \ .
\label{a2}
\end{eqnarray}
The angular integral can be infered by inserting the Gaussian probability 
density~(\ref{gauss}) and considering $\nu=0$, i.e.~$M_0^{(T\beta)}=1$.
The radial integral can then be done and we find as an interesting side result
\begin{eqnarray}
 \int |\Delta_N(\vec{e})|^\beta d\Omega
 =  \frac{\pi^{N/2}\prod_{n=1}^N\Gamma(1+n\beta/2)}
            {2^{\beta N(N-1)/4-1}\Gamma^N(1+\beta/2)\Gamma(\mu/2)} \ .
\label{a3}
\end{eqnarray}
Putting $u=r^2$ and collecting everything we arrive at the second of
Eqs.~(\ref{mome}).

\section{Evaluation of the Kernel}
\label{AppB}

We start by observing that the determinant can be written in the form
\begin{eqnarray}
\prod_{n=1}^N \frac{is_2-H_{0n}}{s_1^--H_{0n}} &=&
\prod_{n=1}^N \left(1+\frac{is_2-s_1}{s_1^--H_{0n}}\right)
                                       \nonumber\\
&=& 1 + \sum_{n=1}^N \frac{is_2-s_1}{s_1^--H_{0n}}
              \sum_{m=0}^{N-1} \frac{(is_2-s_1)^m}
                                                {\prod_{m'\in\Omega_{n,m}^{(N)}}(H_{0n}-H_{0m'})} \ .
\label{b1}
\end{eqnarray}
Here, $\Omega_{n,m}^{(N)}$ is a subset of the $N-1$ indices remaining
when removing the index $n$ from the original $N$ indices such that
$m$ pairs $(n,m')$ are formed. This can also be formulated in terms of
symmetric functions. For example, in the case $N=3$ and $n=1$, one has
\begin{eqnarray}
& & \sum_{m=0}^{2} \frac{(is_2-s_1)^m}
                                                {\prod_{m'\in\Omega_{1,m}^{(3)}}(H_{0n}-H_{0m'})} 
                              \nonumber\\
& & \qquad = 1 +  \frac{is_2-s_1}{H_{01}-H_{02}} +  \frac{is_2-s_1}{H_{01}-H_{03}}
        + \frac{(is_2-s_1)^2}{(H_{01}-H_{02})(H_{01}-H_{03})} \ .
\label{b2}
\end{eqnarray}
This yields a most convenient expression containing only $\delta$ functions,
\begin{eqnarray}
\Im\prod_{n=1}^N \frac{is_2-H_{0n}}{s_1^--H_{0n}}  &=&
\pi \sum_{n=1}^N (is_2-s_1)\delta(s_1-H_{0n})
                        \nonumber\\
& & \qquad\qquad\qquad     \sum_{m=0}^{N-1} \frac{(is_2-s_1)^m}
                                                {\prod_{m'\in\Omega_{n,m}^{(N)}}(H_{0n}-H_{0m'})} \ ,
\label{b3}
\end{eqnarray}
which facilitates the evaluation of the $s_1$ integral in Eq.~(\ref{gueker}). Importantly,
the difference $s_1-is_2$ also disappears in the denominator and the remaining
$s_2$ integration simply becomes 
\begin{eqnarray}
& & \int\limits_{-\infty}^{+\infty} \left(is_2-H_{0n}\right)^m \, 
                                            \exp\left(\frac{(is_2-x_q)^2}{2t\alpha^2}\right) ds_2
                                               \nonumber\\
& & \qquad\qquad\qquad\qquad =  \frac{\sqrt{2t\alpha^2}^{m+1}\sqrt{\pi}}{2^m} \, 
                          \mathbf{H}_m\left(\frac{x_q-H_{0n}}{\sqrt{2t\alpha^2}}\right) \ ,
\label{b4}
\end{eqnarray}
with $\mathbf{H}_m(z)$ denoting the Hermite polynomial of degree $m$. Collecting
everything, we find
\begin{eqnarray}
C_N(x_p,x_q,t\alpha^2,H_0) &=& \sum_{m=0}^{N-1} \left(\frac{t\alpha^2}{2}\right)^{m/2}
          \sum_{n=1}^N  \frac{\exp\left((H_{0n}-x_p)^2/2t\alpha^2\right)}{\sqrt{2\pi t\alpha^2}}
                                                           \nonumber\\
& & \qquad\qquad\qquad\qquad \frac{\mathbf{H}_m((x_q-H_{0n})/\sqrt{2t\alpha^2})}
                                           {\prod_{m'\in\Omega_{n,m}^{(N)}}(H_{0n}-H_{0m'})} \ .
\label{b5}
\end{eqnarray}
We notice that the first sum extends from zero to $N-1$, exactly as in the case of the
GUE without external field.

\end{document}